\documentclass[12pt]{elsarticle}                                          
\usepackage{axodraw4j}
\usepackage{hyperref}
\usepackage{graphicx}                                                         
\usepackage{a41}                                                                
\usepackage{xcolor}                     
\usepackage[rflt]{floatflt}
\usepackage{float}
\usepackage{lscape}
\usepackage{hyperref}
\hypersetup{colorlinks=true, 
linkcolor=blue, filecolor=blue, urlcolor=mygreen}
\usepackage{breakurl} 
\usepackage{times}     %
\usepackage{array}
\usepackage[english]{babel}
\usepackage[latin1]{inputenc}
\usepackage[T1]{fontenc}
\usepackage{ae}
\usepackage{url}
\usepackage{amsmath, amsthm, amssymb}
\usepackage{slashed}

\usepackage{rotating}
\usepackage{graphicx}
\usepackage{comment}
\newcounter{mmacnt}
\def\restartmma{\setcounter{mmacnt}{0}}
\restartmma \catcode`|=\active
\def|#1|{\mathrm{#1}}
\catcode`|=12
\newenvironment{mma}{
\par\smallskip
\catcode`|=\active
\parskip=0pt\parindent=0pt 
\small
\def\In##1\\{%
\def\linebreak{\hfill\break\null\qquad}%
\refstepcounter{mmacnt}
\hangindent=2.5em\hangafter=0
\leavevmode
\llap{\tiny\sffamily In[\arabic{mmacnt}]:=\kern.5em}%
\mathversion{bold}\footnotesize$
\displaystyle##1$\normalsize
\mathversion{normal}\par
 }%
\def\Print##1\\{%
\def\linebreak{\hfill\break}%
\hangindent=2.5em\hangafter=0
\leavevmode ##1\par}%
\def\Out##1\\{%
\def\linebreak{$\hfill\break\null\hfill$}%
\kern\abovedisplayskip\par
\hangindent=2.5em\hangafter=0
\leavevmode
\llap{\tiny\sffamily Out[\arabic{mmacnt}]=\kern.5em}
\footnotesize$\displaystyle##1$
\normalsize\hfill\null\par
\kern\belowdisplayskip
}%
\def\Warning##1##2\\{%
\def\linebreak{\hfill\break}%
\hangindent=2.5em\hangafter=0
\leavevmode
{\scriptsize##1 : ##2}\par}%
}{%
\par\smallskip
}

\usepackage{color}
\newenvironment{fshaded}{%
\MakeFramed {\FrameRestore}
}%
{\endMakeFramed}

\makeatletter
\def\ps@pprintTitle{%
\let\@oddhead\@empty
\let\@evenhead\@empty
\def\@oddfoot{\reset@font\hfil\thepage\hfil}
\let\@evenfoot\@oddfoot
}
\makeatother
\usepackage{tikz}
\usetikzlibrary{matrix}
\allowdisplaybreaks[4]
\newcommand{\n}{\nonumber}
\begin{document}
\begin{frontmatter}
\title{\Large
\textbf{General one-loop contributions to the decay 
$H\rightarrow \nu_l\bar{\nu}_l\gamma$}}
\author{Khiem Hong Phan$^{1,2}$}
\author{Dzung Tri Tran$^{3}$}
\author{and Le Tho Hue$^{4}$}
\address{\it $^{1)}$Institute of Fundamental and Applied Sciences, 
Duy Tan University, Ho Chi Minh City $700000$, Vietnam\\ 
$^{2)}$Faculty of Natural Sciences, Duy Tan University, 
Da Nang City $550000$, Vietnam\\
\it 
$^{3)}$University of Science Ho Chi Minh City, $227$ 
Nguyen Van Cu, District $5$, HCM City, Vietnam\\
\it $^{4)}$Institute of Physics, Vietnam Academy 
of Science and Technology, 
$10$ Dao Tan, Ba Dinh, Hanoi, Vietnam}
\pagestyle{myheadings}
\markright{}
\begin{abstract} 
General one-loop contributions to the decay  amplitudes 
$H\rightarrow \nu_l\bar{\nu}_l\gamma$  are presented, 
considering all possible contributions of  additional
heavy vector gauge bosons, fermions, and charged 
(and also neutral) scalar particles appearing in the loop 
diagrams. Moreover, the results can be applied directly 
when extra neutrinos (apart from three ones  in 
standard model) are taken into account
in final states. Analytic results are presented 
in terms of Passarino-Veltman scalar functions which can 
be evaluated numerically using {\tt LoopTools}.  
In the standard model framework,  these analytical results 
are generated and cross-checked with previous 
computations.  We find that our results  are well consistent 
with these computations. Within standard model limit, 
phenomenological results for the decay channels  are  
also studied using the updated input parameters
at the Large Hadron Collider.  
\end{abstract}
\end{frontmatter}
\section{Introduction}
\noindent
Searching for all decay modes of the standard model-like 
(SM-like) Higgs boson are one of the main purposes
at the High Luminosity Large Hadron Collider (HL-LHC) 
\cite{ATLAS:2013hta,CMS:2013xfa} as well as Future 
Lepton Colliders~\cite{Baer:2013cma}. Because 
the partial decay widths of Higgs boson 
contain an important information for testing 
the nature of Higgs sector. Among the Higgs decay modes, 
the channels of $H\rightarrow$ invisible particles 
\cite{Sirunyan:2018owy,Aaboud:2019rtt,Aaboud:2018sfi,
Ngairangbam:2020ksz,Aad:2012re,Belanger:2013kya,Heikinheimo:2012yd}
and $H\rightarrow \gamma$ plus invisible particles
\cite{Sirunyan:2019xst, Sirunyan:2020zow} 
are great of interest by following reasons.  First, 
these decay processes can be measured at the LHC
\cite{Sirunyan:2018owy,Aaboud:2019rtt,Aaboud:2018sfi,Aad:2012re,
Sirunyan:2019xst, Sirunyan:2020zow}.
Therefore, they could be used for verifying the standard 
model at higher energy regions. On the other hands, 
there are exist many models beyond the standard model (BSM) 
in which new invisible particles 
rather than neutrinos are proposed. In addition, many 
new heavy particles that are absent in SM may exchange in 
the loop diagrams of the aforementioned decay channels. 
The leading contributions to the decays 
$H\rightarrow \nu_l\bar{\nu}_l\gamma$ are from one-loop level 
without  diagrams containing photon exchanges. 
The decay rates will be more sensitive with new 
contributions from BSMs than  those of the decays 
$H\to  e^+e^-\gamma, \mu^+\mu^-\gamma$, which consist of  
both tree contributions from Yukawa coupling $Hl\bar{l}$ and 
one loop contributions from photon 
exchange~\cite{Chen:2012ju, Dicus:2013ycd,  Kachanovich:2020xyg}. 
With the latest experimental report, evidence of  these 
two decay channels have been concerned at $3.2\sigma$ over 
background and $m_{ll}<30$~GeV~\cite{Aad:2021jqf}. 
Correspondingly, the contributions from $Z$ exchanges 
do not appear in this observation, but CMS collaboration 
has been searching for them through the channels
$H\to Z\gamma \to l\bar{l} \gamma$~\cite{CMS:2018bzq}, 
which may have the same properties as the  decays  
$H\rightarrow \nu_l\bar{\nu}_l\gamma$ in the SM 
framework. As a result, the decay widths of 
$H\rightarrow \nu_l\bar{\nu}_l\gamma$ could 
provide an useful tool for controlling SM background 
as well as constraining new physics parameters.

One-loop formulas for $H\rightarrow \nu_l\bar{\nu}_l \gamma$ 
within SM framework have computed in Ref.~\cite{Sun:2013cba}. 
Besides that, a model 
independent for investigating Higgs decay to a 
photon and invisible particles has been proposed in 
Ref.~\cite{Kamenik:2012hn}. 
The decay channel of Higgs to a photon and light vector 
gauge bosons which they belong to $U(1)$ extension of SM 
has also considered in Ref.~\cite{Davoudiasl:2013aya}. 
In next-to-minimal supersymmetry (SUSY) framework, Higgs decay to photon 
plus pair of lightest susy particles have studied 
in Ref.~\cite{Curtin:2013fra}. In Ref.~\cite{Curtin:2013fra}, 
the decay process has been used for 
probing dark matter as well as constraining SUSY parameters. 
Supersymmetry-breaking scale has been examined through 
the Higgs decay to photon and gravitinos in 
\cite{Petersson:2012dp}.

In this article, we present general one-loop formulas 
for the decay $H\rightarrow \nu_l\bar{\nu}_l\gamma$. 
The results are valid for many BSMs 
which new heavy vector bosons, fermions, and  scalar 
particles predicted by these models are considered in 
the loop diagrams. Moreover, the calculations can 
be extended directly
when the extra neutrinos (rather than $3$ in 
standard model) are taken into account
in final states. Analytic results are presented 
in terms of Passarino-Veltman scalar 
functions which can be computed 
numerically by using the package {\tt LoopTools}. 
The calculations are also verified numerically by 
checking the ultraviolet finiteness of the results. 
We find that the results are good stability when 
varying ultraviolet cutoff parameters. 
The results are then applied to the case of standard model 
which the decay rates are generated and cross-checked with 
the previous computations. Our results in this work 
are good agreement with the previous references. 
All physical results for the decay channels within SM 
are examined with taking updated input parameters at the 
Large Hadron Collider. While the phenomenological 
results for the decay processes in several BSMs 
are referred to our next papers.

The results of this work can be applied calculate one-loop contributions of new particles predicted by well-known BSM constructed previously, for examples  many popular SM extensions including only new charged scalars such as  two Higgs doublet models. In the SUSY model, 
new loop contributions come from charged Higgs bosons,  superpartners of leptons and  gauge bosons. One loop contribution from new charged gauge bosons may appear in many electroweak gauge extensions 
such as the left-right models (LR) constructed 
from the $SU(2)_L\times SU(2)_R\times U(1)_Y$~\cite{Pati:1974yy, Mohapatra:1974gc, Senjanovic:1975rk},  the 3-3-1 models ($SU(3)_L\times U(1)_X$)~\cite{Singer:1980sw, Valle:1983dk, Pisano:1991ee, Frampton:1992wt, Diaz:2004fs,Fonseca:2016tbn, Foot:1994ym}, the  $3$-$4$-$1$ models ($SU(4)_L\times U(1)_X$)~\cite{Foot:1994ym, Sanchez:2004uf, Ponce:2006vw, Riazuddin:2008yx, Jaramillo:2011qu, Long:2016lmj}, ect.  These one-loop contributions may be significant in the amplitudes of  the mentioned decay processes. Phenomenological results for
the decay processes in the mentions models will be very interesting for 
further studies, which will be our future projects.

The layout of the paper is as follows: In section 2,
we present briefly one-loop tensor reduction method. 
Detailed calculations for one-loop 
contributions to $H\rightarrow \nu_l\bar{\nu}_l \gamma$ are presented 
in this section. Conclusions and outlook are devoted in section $3$. 
In appendices, Feynman rules and involving couplings in the 
decay processes are shown. 
\section{Calculation}    
Detailed calculations for one-loop 
contributions to $H\rightarrow \nu_l\bar{\nu}_l \gamma$ are presented 
in this section. We first describe briefly one-loop tensor reduction
method in the following subsection. General analytic results and 
physical results of the decay processes are then 
shown in the next subsections. 
\subsection{Method} 
\noindent
In this calculation, we follow tensor reduction method developed in 
Ref.~\cite{Denner:2005nn}. Following the technique, tensor one-loop 
integrals with $N$-external lines can be decomposed into scalar 
functions with $N\leq 4$. The approach will be explained briefly in 
the following paragraphs.

First, one-loop one-, two-, three- and four-point tensor integrals 
with rank $P$ are defined:
\begin{eqnarray}
\{A; B; C; D\}^{\mu_1\mu_2\cdots \mu_P}= (\mu^2)^{2-d/2}
\int \frac{d^dk}{(2\pi)^d} 
\dfrac{k^{\mu_1}k^{\mu_2}\cdots k^{\mu_P}}{\{D_1; D_1 D_2;D_1D_2D_3; 
D_1D_2D_3D_4\}}.
\end{eqnarray}
In this formula, 
$D_j$ ($j=1,\cdots, 4$) are the inverse Feynman propagators
\begin{eqnarray}
 D_j = (k+ q_j)^2 -m_j^2 +i\rho,
\end{eqnarray}
$q_j = \sum\limits_{i=1}^j p_i$, 
$p_i$ are the external momenta,  and  
$m_j$ are internal masses in the loops.
We are working on space-time dimension 
$d=4-2\varepsilon$. The parameter $\mu^2$ 
plays role of a renormalization
scale. We then present explicit reduction 
formulas for one-loop one-, two-, three- and four-point tensor
integrals up to rank $P=3$ as follows~\cite{Denner:2005nn}:
\begin{eqnarray}
A^{\mu}        &=& 0, \\
A^{\mu\nu}     &=& g^{\mu\nu} A_{00}, \\
A^{\mu\nu\rho} &=& 0,\\
B^{\mu}        &=& q^{\mu} B_1,\\
B^{\mu\nu}     &=& g^{\mu\nu} B_{00} + q^{\mu}q^{\nu} B_{11}, \\
B^{\mu\nu\rho} &=& \{g, q\}^{\mu\nu\rho} B_{001} 
+ q^{\mu}q^{\nu}q^{\rho} B_{111}, \\
C^{\mu}        &=& q_1^{\mu} C_1 + q_2^{\mu} C_2 
 = \sum\limits_{i=1}^2q_i^{\mu} C_i, 
\\
C^{\mu\nu}    &=& g^{\mu\nu} C_{00} 
 + \sum\limits_{i,j=1}^2q_i^{\mu}q_j^{\nu} C_{ij},
\\
C^{\mu\nu\rho} &=&
\sum_{i=1}^2 \{g,q_i\}^{\mu\nu\rho} C_{00i}+
\sum_{i,j,k=1}^2 q^{\mu}_i q^{\nu}_j q^{\rho}_k C_{ijk},\\
D^{\mu}        &=& q_1^{\mu} D_1 + q_2^{\mu} D_2 + q_3^{\mu}D_3 
 = \sum\limits_{i=1}^3q_i^{\mu} D_i, \\
 D^{\mu\nu}    &=& g^{\mu\nu} D_{00} 
 + \sum\limits_{i,j=1}^3q_i^{\mu}q_j^{\nu} D_{ij}, 
\\
D^{\mu\nu\rho} &=&
	\sum_{i=1}^3 \{g,q_i\}^{\mu\nu\rho} D_{00i}+
	\sum_{i,j,k=1}^3 q^{\mu}_i q^{\nu}_j q^{\rho}_k D_{ijk}.
\end{eqnarray}
The short notation~\cite{Denner:2005nn} 
$\{g, q_i\}^{\mu\nu\rho}$ is used 
as follows: $\{g, q_i\}^{\mu\nu\rho} = g^{\mu\nu} q^{\rho}_i 
+ g^{\nu\rho} q^{\mu}_i + g^{\mu\rho} q^{\nu}_i$ in the 
above relations. The scalar coefficients $A_{00}, B_1, \cdots, D_{333}$
in the right hand sides of the above equations 
are so-called Passarino-Veltman functions (PV)~\cite{Denner:2005nn}. 
Analytic formulas of the PV functions are well-known 
and they have been implemented into 
{\tt LoopTools}~\cite{Hahn:1998yk} for 
numerical computations. 
\subsection{General one-loop contributions 
to $H\rightarrow \nu_l\bar{\nu}_l\gamma$} 
General one-loop contributions
to $H (p) \rightarrow \nu_l (q_1)
\bar{\nu}_l (q_2)\gamma (q_3)$ in 
arbitrary beyond the standard models are 
calculated in this section. One-loop Feynman
diagrams involving the decay processes can be 
grouped into several classes shown in the 
following paragraphs. For on-shell external 
photon, the ward identity is implied. 
As a result, we apply the following relation:
$q_3^{\nu}\epsilon_{\nu}^*=0$ where $q_3^{\nu}$,
$\epsilon_{\nu}^*$ are momentum and polarization 
vector of the external photon respectively. 
Kinematic invariant variables 
involving to the decay processes are included: 
\begin{eqnarray}
 p^2 &=& M_H^2, \quad q_1^2= q_2^2=q_3^2=0, \nonumber\\
 q_{12}&=&q^2=(q_1+q_2)^2=2 q_1\cdot q_2,\quad  
 q_{13} = 2q_1\cdot q_3,\quad
 q_{23} = 2q_2\cdot q_3.
\end{eqnarray}
The general one-loop amplitude 
which obeys the invariant Lorentz structure
can be decomposed as follows~\cite{Kachanovich:2020xyg}:
\begin{eqnarray}
\label{FLR-amp}
 \mathcal{A}_{\text{loop}} &=&\sum\limits_{k=1}^2 
 \Big\{
[q_3^{\mu}q_i^{\nu} - g^{\mu\nu}q_3\cdot q_k] \bar{u}(q_1)
(F_{k,R} \gamma_{\mu}P_R + F_{k,L} \gamma_{\mu}P_L) v(q_2) 
\Big\}\varepsilon^{*}_{\nu}.
\end{eqnarray}
In this equation, all form factors are computed as follows:
\begin{eqnarray}
 F_{k,L/R} &=& F_{k,L/R}^{\text{Trig}} + F_{k,L/R}^{\text{Box} }
\end{eqnarray}
for $k=1,2$. Each form factor in (\ref{FLR-amp}) will 
be contributed from different kind of particles 
such as vector bosons $V_i$, charged scalar particles 
$S_i$ and fermions $f_i$ exchanging in loop diagrams. 
These particles appear in many BSMs whose the Feynman 
rules are collected in 
Tables \ref{Feynman rules table} and~\ref{couplings table}. 
After using them to write down all one-loop 
contributions to the  decay amplitudes, the
{\tt Package-X}~\cite{Patel:2015tea} will be used
to  contract  all Dirac traces in the general 
dimension $d$. The analytic formulas of all 
one-loop contributions will be then decomposed 
into one-loop tensor integrals. In this step, 
the above tensor reduction method
is employed to transform  all tensor integrals
into scalar functions included in the  form factors 
$F_{k,L/R}$. Finally, they are collected as 
functions of the well-known  
Passarino-Veltman scalar 
coefficients~\cite{Denner:2005nn,Hahn:1998yk}.
\subsubsection{One-loop triangle diagrams} 
\begin{center}
\begin{figure}[ht]
\hspace{1.5cm}
\includegraphics[width=14cm, height=8cm]
{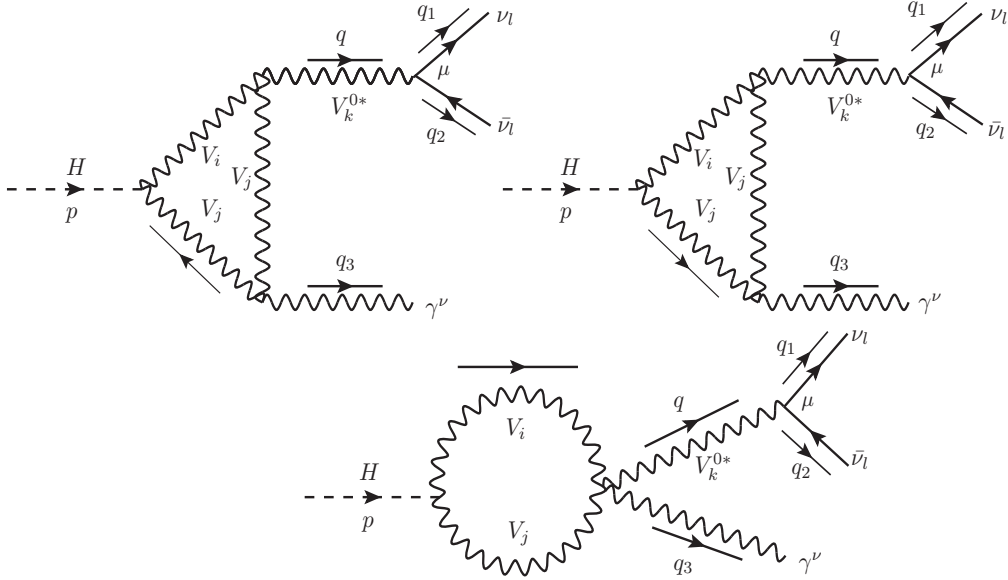}
\caption{\label{ViVJ-dig} One-loop triangle diagrams with 
exchanging vector bosons $V_{i,j}$ particles in loop.}
\end{figure}
\end{center}

We are going to present the calculation in detail.
We first arrive at 
the contributions of one-loop triangle diagrams 
with exchanging vector bosons $V_i, V_j$ in loop 
(seen Fig.~\ref{ViVJ-dig}). 
By applying one-loop tensor reduction method in the 
previous subsection, the form factors are expressed 
in terms of PV-functions as follows: 
\begin{eqnarray}
\label{FLR-ViVj}
F_{k,L}^{\text{Trig}}|_{V_i,V_j} 
&=& \sum\limits_{V_i,V_j}
\dfrac{g_{HViVj} \; 
g_{V^0_k \nu_l \bar{\nu}_l}^{L} }{32 \pi ^2 M_{V_i}^2 M_{V_j}^2
(q^2 - M_{V^0_k}^2 + i\Gamma_{V^0_k}M_{V^0_k} )}
\times
\n \\
&&\hspace{0cm} \times
\Bigg\{
\Big[
eQ  g_{V^0_k V_i V_j} (M_H^2 + M_{V_i}^2 + M_{V_j}^2)
+2 g_{V^0_k A V_i V_j} (M_H^2 - M_{V_i}^2)
\Big]
B_{11}(M_H^2,M_{V_j}^2,M_{V_i}^2)
\n \\
&&\hspace{0cm}
+
\Big[
eQ  g_{V^0_k V_i V_j} (M_H^2 - M_{V_i}^2 + 3 M_{V_j}^2)
-2 g_{V^0_k A V_i V_j} M_{V_j}^2
\Big]
B_1(M_H^2,M_{V_j}^2,M_{V_i}^2) 
\n \\
&&\hspace{-0cm}
+2 M_{V_j}^2
\Big[
eQ  g_{V^0_k V_i V_j}
-g_{V^0_k A V_i V_j}
\Big] 
B_0(M_H^2,M_{V_j}^2,M_{V_i}^2)
 \\
&&\hspace{-0cm}
+2 g_{V^0_k A V_i V_j} 
\Big[
M_H^2 B_{111}
+ B_{00}
+2 B_{001}
\Big] (M_H^2,M_{V_j}^2,M_{V_i}^2)
\n \\
&&\hspace{0cm}
+4 eQ  g_{V^0_k V_i V_j} M_{V_j}^2
\Big(
3 M_{V_i}^2
+ M_{V_j}^2
- q^2
\Big)
C_0(0,q^2,M_H^2,M_{V_j}^2,M_{V_j}^2,M_{V_i}^2)
\n \\
&&\hspace{0cm}
+ 2 eQ  g_{V^0_k V_i V_j} \Big[
M_H^2 (M_{V_i}^2
+ M_{V_j}^2 
- q^2)
+(4 d - 6) M_{V_i}^2 M_{V_j}^2  
+ M_{V_i}^4 
+ M_{V_j}^4 
\n\\
&&
\hspace{3.5cm}
- q^2 (M_{V_i}^2 + M_{V_j}^2)
\Big]
(C_{22}+C_{12})(0,q^2,M_H^2,M_{V_j}^2,M_{V_j}^2,M_{V_i}^2)
\n \\
&&\hspace{-0cm}
+ 2 eQ  g_{V^0_k V_i V_j}
\Big[
 M_H^2 (M_{V_i}^2 
+  M_{V_j}^2 
- q^2 )
+(4 d - 6) M_{V_i}^2 M_{V_j}^2 
+3  M_{V_j}^4 
\n\\
&&
\hspace{3.5cm}
- M_{V_i}^4 
+ q^2 ( M_{V_i}^2 
-3  M_{V_j}^2 )
\Big]
C_2(0,q^2,M_H^2,M_{V_j}^2,M_{V_j}^2,M_{V_i}^2)
 \Bigg\}, 
 \n\\
 \label{FRL-ViVj}
F_{k,R}^{\text{Trig}}|_{V_i,V_j}
&=&F_{k,L}^{\text{Trig}}|_{V_i,V_j}
(
g_{V^0_k \nu_l \bar{\nu}_l}^{L} \rightarrow
g_{V^0_k \nu_l \bar{\nu}_l}^{R}
). 
\end{eqnarray}
We note that the form factors 
follow the relation: $F_{k,L/R}^{\text{Trig}}
=F_{1,L/R}^{\text{Trig}}
=F_{2,L/R}^{\text{Trig}}$ and 
$F_{k,R}^{\text{Trig}}|_{V_i,V_j}$ can be obtained
directly by replacing $g_{V^0_k \nu_l \bar{\nu}_l}^{L} \rightarrow
g_{V^0_k \nu_l \bar{\nu}_l}^{R}$ in $F_{k,L}^{\text{Trig}}|_{V_i,V_j}$
(as shown in Eq.~(\ref{FRL-ViVj})).

\begin{center}
\begin{figure}[ht]
\hspace{1.5cm}
\includegraphics[width=14.0cm, height=5.0cm]
{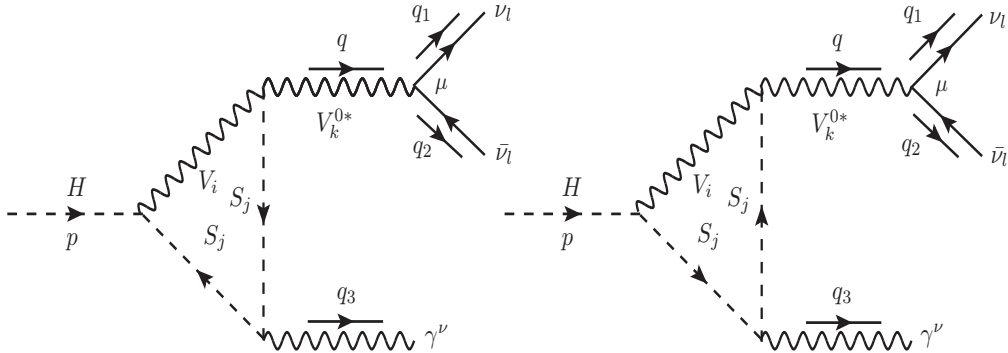}
\caption{\label{ViSj-dig} One-loop triangle 
diagrams with an vector boson $V_i$
and two scalar bosons $S_{j}$ exchanging in loop.}
\end{figure}
\end{center}

\begin{center}
\begin{figure}[ht]
\hspace{1.5cm}
\includegraphics[width=14.0cm, height=6.0cm]
{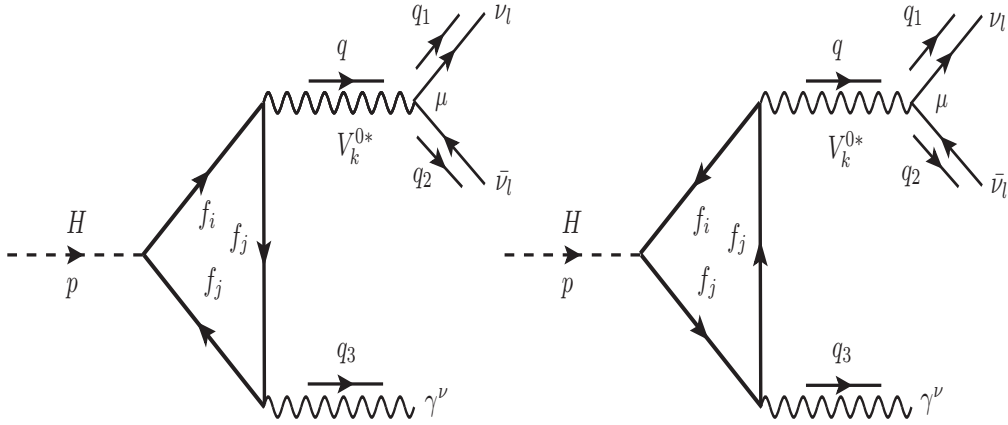}
\caption{\label{ff-dig} Feynman triangle diagrams 
with fermion $f_{i/j}$ particles exchanging in loop.}
\end{figure}
\end{center}

We next take into account the attributions of
one-loop triangle graphs which a boson $V_i$ 
and two charged scalar particles $S_j$ 
are internal lines (as shown in Fig.~\ref{ViSj-dig}).  
Applying the same procedure, the form factors
read:
\begin{eqnarray}
\label{FLR_ViSj}
F_{k,L}^{\text{Trig}}|_{V_i,S_j}
&=& \sum\limits_{Vi,Sj}
\dfrac{eQ  g_{HV_iS_j} \; g_{V^0_k V_iS_j}
\; g_{V^0_k \nu_l \bar{\nu}_l}^{L} 
}{8 \pi ^2 M_{V_i}^2 
(q^2 - M_{V^0_k}^2 + i\Gamma_{V^0_k}M_{V^0_k})}
\times
\\
&& \times
\Bigg\{
(M_{S_j}^2-M_{V_i}^2-M_H^2) 
\Big[
C_{22}+C_{12}
\Big]
(0,q^2,M_H^2,M_{S_j}^2,M_{S_j}^2,M_{V_i}^2)
\n \\
&&\hspace{2cm}
+(M_{S_j}^2+M_{V_i}^2-M_H^2) 
\; C_2(0,q^2,M_H^2,M_{S_j}^2,M_{S_j}^2,M_{V_i}^2)
\Bigg\},
\n\\
F_{k,R}^{\text{Trig}}|_{V_i,S_j} &=&
F_{k,L}^{\text{Trig}}|_{V_i,S_j}
(
g_{V^0_k \nu_l \bar{\nu}_l}^{L} \rightarrow
g_{V^0_k \nu_l \bar{\nu}_l}^R
).
\end{eqnarray}
In addition, we have two vector bosons $V_j$
and a charged scalar $S_i$ exchanging in one-loop 
triangle diagrams (as described as in Fig.~\ref{VjSi-dig}). 
In the same manner as above procedure, 
the form factors $F_{k,L/R}^{\text{Trig}}$ 
are presented as functions of PV-coefficients:
\begin{center}
\begin{figure}[ht]
\hspace{1.5cm}
\includegraphics[width=14.0cm, height=5.0cm]
{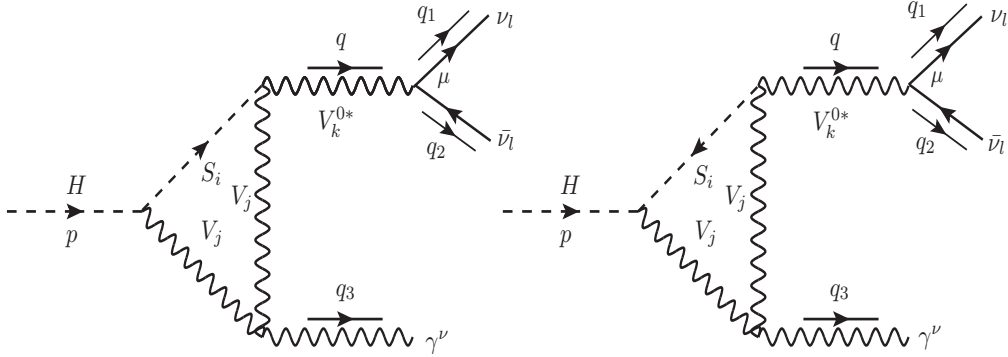}
\caption{\label{VjSi-dig} One-loop triangle diagrams with two 
vector bosons $V_j$ and a scalar boson $S_{i}$ in loop.}
\end{figure}
\end{center}
\begin{eqnarray}
\label{FLR-SiVj}
F_{k,L}^{\text{Trig}}|_{S_i,V_j} 
&=& \sum\limits_{S_i,V_j}
\dfrac{eQ  g_{HS_iV_j} \; g_{V^0_k S_iV_j}
\; g_{V^0_k \nu_l \bar{\nu}_l}^{L} 
}{16 \pi ^2 M_{V_j}^2 
(q^2 - M_{V^0_k}^2 + i\Gamma_{V^0_k}M_{V^0_k}) }
\times \\
&&\hspace{0cm} \times
\Bigg\{
(M_H^2-M_{S_i}^2+M_{V_j}^2) 
\Big[
C_{22}+C_{12}+C_2
\Big](0,q^2,M_H^2,M_{V_j}^2,M_{V_j}^2,M_{S_i}^2)
\n \\
&&\hspace{4cm}
+2M_{V_j}^2 \Big[
C_2+C_0\Big](0,q^2,M_H^2,M_{V_j}^2,M_{V_j}^2,M_{S_i}^2)
\Bigg\},  \n 
\\
F_{k,R}^{\text{Trig}}|_{S_i,V_j} 
&=&F_{k,L}^{\text{Trig}}|_{S_i,V_j} 
(
g_{V^0_k \nu_l \bar{\nu}_l}^{L} \rightarrow
g_{V^0_k \nu_l \bar{\nu}_l}^{R} 
).
\end{eqnarray}
\begin{center}
\begin{figure}[ht]
\hspace{1.5cm}
\includegraphics[width=14.0cm, height=8.0cm]
{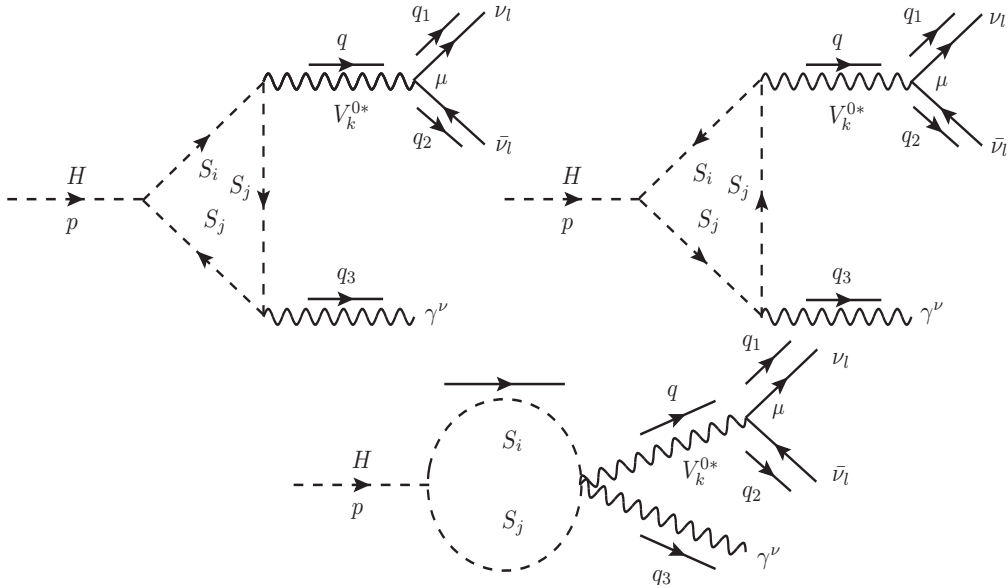}
\caption{\label{SiSj-dig} One-loop bubble and triangle diagrams 
with all charged scalar bosons $S_{i,j}$ internal lines.}
\end{figure}
\end{center}
In further, we also mention 
the attributions  of one-loop bubble 
and triangle diagrams with 
both charged scalar bosons $S_i, S_j$ 
in loop (as depicted in Fig.~\ref{SiSj-dig}).
The resulting for the form factors 
$F_{k,L/R}^{\text{Trig}}$ read 
\begin{eqnarray}
\label{FLR-SiSj}
F_{k,L}^{\text{Trig}}|_{S_i,S_j}
&=& \sum\limits_{S_i, S_j}
\dfrac{eQ  g_{HS_iS_j} \; g_{V^0_k S_iS_j}
\; g_{V^0_k \nu_l \bar{\nu}_l L} 
}{4 \pi ^2 
(q^2 - M_{V^0_k}^2 + i\Gamma_{V^0_k}M_{V^0_k})}
\Big[
C_{22}+C_{12}+C_2
\Big](0,q^2,M_H^2,M_{S_j}^2,M_{S_j}^2,M_{S_i}^2)
,\n
\\
&& \\
F_{k,R}^{\text{Trig}}|_{S_i,S_j} 
&=&F_{k,L}^{\text{Trig}}|_{S_i,S_j}
(
g_{V^0_k \nu_l \bar{\nu}_l}^L \rightarrow
g_{V^0_k \nu_l \bar{\nu}_l}^R 
).
\end{eqnarray}
Lastly, we also have 
fermions exchanging in the loop of the triangle
Feynman diagrams wich are depicted as in 
Fig.~\ref{ff-dig}.
The form factors 
$F_{k,L/R}^{\text{Trig}}$ 
for fermion $f_{i/j}$ contributions can be 
expressed as follows:
\begin{eqnarray}
\label{FLR-fij}
F_{k,L}^{\text{Trig}}|_{f_i,f_j}
&=& \sum\limits_{f_i, f_j}
\dfrac{e Q_f N_C^f
\; g_{V^0_k \nu_l \bar{\nu}_l}^{L}
}{16 \pi ^2 
(q^2 - M_{V^0_k}^2 + i\Gamma_{V^0_k}M_{V^0_k}) }
\; 
(g_{Hf_if_jL} + g_{Hf_if_jR}) 
(g_{V^0_k f_if_jL} + g_{V^0_k f_if_jR})
\times
\n \\
&&\hspace{0cm}
\times
\Bigg\{ 2(m_{f_i}+ m_{f_j}) 
\Big[C_{22}+C_{12}\Big]
(0,q^2,M_H^2,m_{f_j}^2,m_{f_j}^2,m_{f_i}^2)
\\
&&\hspace{.cm}
+(m_{f_i}+3 m_{f_j}) 
C_2(0,q^2,M_H^2,m_{f_j}^2,m_{f_j}^2,m_{f_i}^2)
+m_{f_j} 
C_0(0,q^2,M_H^2,m_{f_j}^2,m_{f_j}^2,m_{f_i}^2)
\Bigg\} ,
\n \\
F_{k,R}^{\text{Trig}}|_{f_i,f_j} 
&=&F_{k,L}^{\text{Trig}}|_{f_i,f_j}
(
g_{V^0_k \nu_l \bar{\nu}_l}^L 
\rightarrow
g_{V^0_k \nu_l \bar{\nu}_l}^R 
).
\end{eqnarray}
\subsubsection{One-loop box diagrams} 
\begin{center}
\begin{figure}[h]
\hspace{1.5cm}
\includegraphics[width=14.0cm, height=10.0cm]
{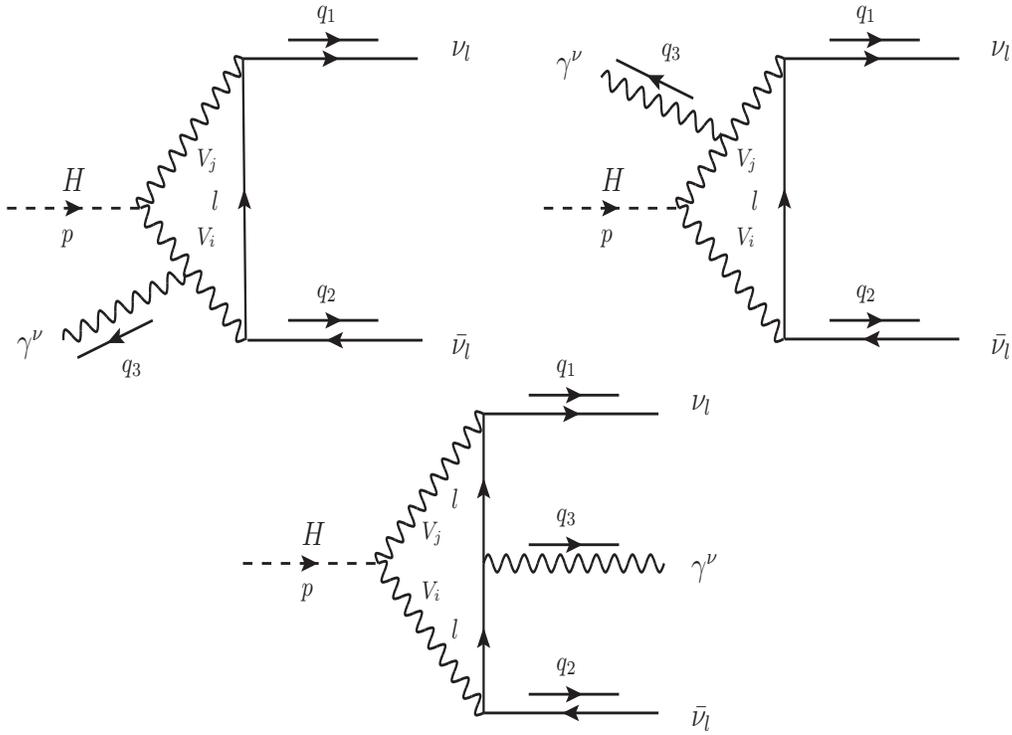} 
\caption{\label{ViVj-box} One-loop box diagrams 
with $V_{i}, V_{j}$ exchanging in the loop.}
\end{figure}
\end{center}

We turn our attention to all one-loop 
box Feynman diagrams contributing to 
the decay processes. Firstly, one-loop four-point 
Feynman diagrams having $V_{i},V_{j}$
in the loop (as described in 
Fig.~\ref{ViVj-box})
are performed.  The form factors 
$F_{k,L/R}^{\text{Box}}$ with $k=1,2$
are then given by 
\begin{eqnarray}
\label{FLRbox-ViVj}
F_{1,L}^{\text{Box}}|_{V_i,V_j}  
&=& \sum\limits_{V_i, V_j}
\dfrac{eQ  g_{HV_iV_j} \; g_{V_i l \nu_l}^{L}  
\; g_{V_j l \nu_l}^{L} 
}{16 \pi ^2 M_{V_i}^2 M_{V_j}^2}
\times
\n\\
&&
\times
\Bigg\{
(M_H^2+M_{V_i}^2+M_{V_j}^2) 
\Big[
(C_{22}+C_{12})(0,q_{12},M_H^2,M_{V_i}^2,M_{V_i}^2,M_{V_j}^2)
\n\\
&&\hspace{5.cm}
+(C_{22}+C_{12})(q_{12},0,M_H^2,M_{V_i}^2,M_{V_j}^2,M_{V_j}^2)
\Big]
\n\\
&&
+(M_H^2+3 M_{V_i}^2-M_{V_j}^2) 
\Big[
C_2(0,q_{12},M_H^2,M_{V_i}^2,M_{V_i}^2,M_{V_j}^2)
\n\\
&&\hspace{5cm}
+C_2(q_{12},0,M_H^2,M_{V_i}^2,M_{V_j}^2,M_{V_j}^2)
\Big]
\n\\
&&
+(2 M_{V_i}^2-2 M_{V_j}^2) 
C_1(q_{12},0,M_H^2,M_{V_i}^2,M_{V_j}^2,M_{V_j}^2)
\n\\
&&
+2 M_{V_i}^2 
\Big[
C_0(0,q_{12},M_H^2,M_{V_i}^2,M_{V_i}^2,M_{V_j}^2)
+ C_0(q_{12},0,M_H^2,M_{V_i}^2,M_{V_j}^2,M_{V_j}^2)
\Big]
\n\\
&&
+m_l^2 
\Big[
(C_{22}+ C_{12}+ C_2)(0,0,q_{13},m_l^2,m_l^2,M_{V_j}^2)
\n\\
&&\hspace{4cm}
- (C_{22}+ C_{12}+ C_2)(0,0,q_{13},m_l^2,M_{V_j}^2,M_{V_j}^2)
\Big]
\n\\
&&
+\Big[
 m_l^2 (M_H^2
+ M_{V_i}^2
+ M_{V_j}^2)
+(2 d -4) M_{V_i}^2 M_{V_j}^2
\Big]
\times
\\
&&
\times
\Big[
(D_{33}+D_{23})
(0,0,0,M_H^2;q_{12},q_{13},M_{V_i}^2,m_l^2,M_{V_j}^2,M_{V_j}^2)
\n\\
&&
\hspace{1.5cm}
+(D_{33}+D_{23}+D_{13})
(0,0,0,M_H^2;q_{23},q_{12},M_{V_i}^2,M_{V_i}^2,m_l^2,M_{V_j}^2)
\n\\
&&\hspace{1.5cm}
-(D_{33}+D_{23})
(0,0,0,M_H^2;q_{23},q_{13},M_{V_i}^2,m_l^2,m_l^2,M_{V_j}^2)
\Big]
\n\\
&&
+\Big[
m_l^2 (M_H^2 
+3 M_{V_i}^2
- M_{V_j}^2)
+(2 d -8) M_{V_i}^2 M_{V_j}^2
\Big] 
\times
\n\\
&&\hspace{3cm} \times
\Big[
D_3(0,0,0,M_H^2;q_{12},q_{13},M_{V_i}^2,m_l^2,M_{V_j}^2,M_{V_j}^2)
\n\\
&&\hspace{3.5cm}
+D_3(0,0,0,M_H^2;q_{23},q_{12},M_{V_i}^2,M_{V_i}^2,m_l^2,M_{V_j}^2)
\n\\
&&\hspace{3.5cm}
- D_3(0,0,0,M_H^2;q_{23},q_{13},M_{V_i}^2,m_l^2,m_l^2,M_{V_j}^2)
\Big]
\n\\
&&
+\Big[
2 m_l^2 (M_{V_i}^2
- M_{V_j}^2)
-4 M_{V_i}^2 M_{V_j}^2
\Big] 
D_2(0,0,0,M_H^2;q_{12},q_{13},M_{V_i}^2,m_l^2,M_{V_j}^2,M_{V_j}^2)
\n\\
&&
+2 m_l^2 M_{V_i}^2 
\Big[
D_0(0,0,0,M_H^2;q_{12},q_{13},M_{V_i}^2,m_l^2,M_{V_j}^2,M_{V_j}^2)
\n\\
&&\hspace{3cm}
+ D_0(0,0,0,M_H^2;q_{23},q_{12},M_{V_i}^2,M_{V_i}^2,m_l^2,M_{V_j}^2)
\n\\
&&\hspace{3.5cm}
- D_0(0,0,0,M_H^2;q_{23},q_{13},M_{V_i}^2,m_l^2,m_l^2,M_{V_j}^2)
\Big]
\Bigg\},
\n\\
F_{1,R}^{\text{Box}}|_{V_i,V_j}
&=& F_{1,L}^{\text{Box}}|_{V_i,V_j} 
(
g_{V_i l \nu_l}^L \rightarrow
g_{V_i l \nu_l}^R; 
\; g_{V_j l \nu_l}^{L}  \rightarrow
g_{V_j l \nu_l}^{R} 
),\\
F_{2,L}^{\text{Box}}|_{V_i,V_j}
&=& 
F_{1,L}^{\text{Box}}|_{V_i,V_j} 
(
\{q_{13}, q_{23}\}
\rightarrow 
\{q_{23}, q_{13}\}
),
\\
F_{2,R}^{\text{Box}}|_{V_i,V_j} 
&=& F_{2,L}^{\text{Box}}|_{V_i,V_j}
(
g_{V_i l \nu_l}^L 
\rightarrow
g_{V_i l \nu_l}^R; 
\; g_{V_j l \nu_l}^{L}  
\rightarrow
g_{V_j l \nu_l}^{R} 
).
\end{eqnarray}
We find that analytic results for the above form factors
are given up to $D_{33}$-coefficient functions. The reason for that 
fact can be explained as follows. Although tensor one-loop box 
integrals with rank $P\geq 4$ appear in each Feynman diagram in 
Fig.~\ref{ViVj-box}, we find that these terms are cancelled out 
after summing all diagrams. Consequently, the amplitudes are 
only decomposed up to one-loop box integrals with rank $P=2$. 
\begin{center}
\begin{figure}[ht]
\hspace{1.5cm}
\includegraphics[width=14.0cm, height=10.0cm]
{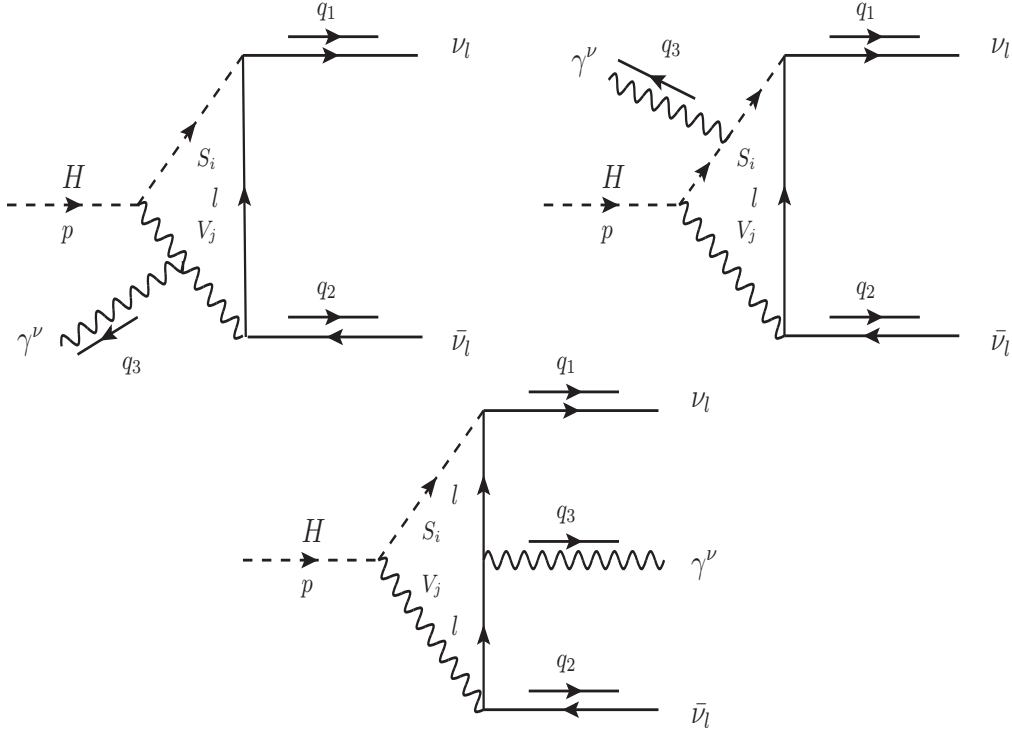}
\caption{\label{VS-box} Feynman box diagrams 
for $V_{i,j}, S_{i,j}$.}
\end{figure}
\end{center}
\begin{center}
\begin{figure}[ht]
\hspace{1.5cm}
\includegraphics[width=14.0cm, height=10.0cm]
{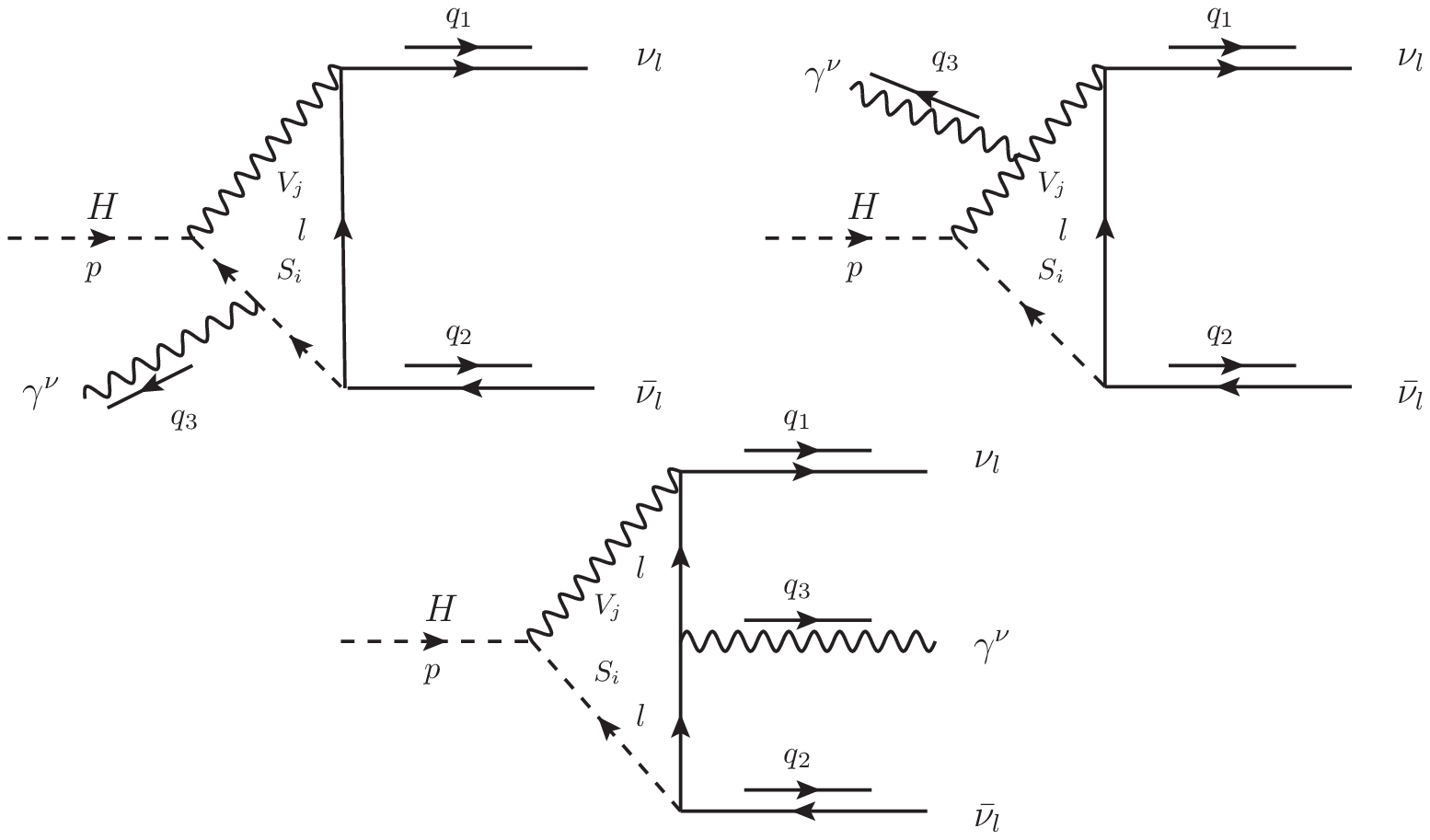}
\caption{\label{SV-box} Feynman box 
diagrams for $V_{i,j}, S_{i,j}$.}
\end{figure}
\end{center}
We next consider one-loop box diagrams 
with  $V_i, S_j$ in the loop.
In order to get the symmetry of 
$F_{k,L/R}^{\text{Box}}$ which 
follow the relation
\begin{eqnarray}
F_{1,L/R}^{\text{Box}}|_{V_i,S_j} 
= F_{2,L/R}^{\text{Box}}|_{V_i,S_j}
(\{q_{13}, q_{23}\} \rightarrow \{q_{23}, q_{13}\}),
\end{eqnarray}
we should consider $6$ diagrams 
as shown in Fig.~\ref{VS-box}
and Fig.~\ref{SV-box} together. This is because 
that the coupling of charged scalar $S_i^{\pm}$ 
to $l\bar{\nu}_l$ (and to $\bar{l}\nu_l$) take
the different forms (seen Table 2 for more detail). 
The form factors 
$F_{k,L/R}^{\text{Box}}$ 
are then  presented  as follows:
\begin{eqnarray}
\label{FLRBox-ViSj}
F_{1,L}^{\text{Box}}|_{V_i,S_j} 
&=& \sum\limits_{V_i, S_j}
\dfrac{eQ  m_l \; g_{HV_iS_j} \; g_{S_j l \nu_l}^R
g_{V_i l \nu_l}^L }{16 \pi ^2 M_{V_i}^2}
\times
\n\\
&&
\times
\Bigg\{
(M_H^2-M_{S_j}^2+M_{V_i}^2) 
\Big[
(D_{33}+D_{23}+D_{13})(0,0,0,M_H^2;q_{23},q_{12},M_{V_i}^2,M_{V_i}^2,m_l^2,M_{S_j}^2)
\n\\
&&\hspace{4cm}
-(D_{33}+D_{23}+D_{13})(0,0,0,M_H^2;q_{23},q_{12},M_{S_j}^2,M_{S_j}^2,m_l^2,M_{V_i}^2)
\n\\
&&\hspace{4cm}
-(D_{33}+D_{23})(0,0,0,M_H^2;q_{12},q_{13},M_{V_i}^2,m_l^2,M_{S_j}^2,M_{S_j}^2)
\n\\
&&\hspace{4cm}
+ (D_{33}+D_{23})(0,0,0,M_H^2;q_{12},q_{13},M_{S_j}^2,m_l^2,M_{V_i}^2,M_{V_i}^2)
\n\\
&&\hspace{4cm}
-(D_{33}+D_{23})(0,0,0,M_H^2;q_{23},q_{13},M_{V_i}^2,m_l^2,m_l^2,M_{S_j}^2)
\n\\
&&\hspace{4cm}
-(D_{33}+D_{23})(0,0,0,M_H^2;q_{23},q_{13},M_{S_j}^2,m_l^2,m_l^2,M_{V_i}^2)
\Big]
\n \\
&&
+(M_H^2-M_{S_j}^2+3 M_{V_i}^2) 
\Big[
D_3(0,0,0,M_H^2;q_{23},q_{12},M_{V_i}^2,M_{V_i}^2,m_l^2,M_{S_j}^2)
\n\\
&&\hspace{4cm}
-D_3(0,0,0,M_H^2;q_{12},q_{13},M_{V_i}^2,m_l^2,M_{S_j}^2,M_{S_j}^2)
\n \\
&&\hspace{4.0cm}
-D_3(0,0,0,M_H^2;q_{23},q_{13},M_{V_i}^2,m_l^2,m_l^2,M_{S_j}^2)
\Big]
\n\\
&&
+(M_H^2-M_{S_j}^2-M_{V_i}^2) 
\Big[
D_3(0,0,0,M_H^2;q_{12},q_{13},M_{S_j}^2,m_l^2,M_{V_i}^2,M_{V_i}^2)
\n \\
&&\hspace{4cm}
-D_3(0,0,0,M_H^2;q_{23},q_{12},M_{S_j}^2,M_{S_j}^2,m_l^2,M_{V_i}^2)
\n\\
&&\hspace{4.0cm}
-D_3(0,0,0,M_H^2;q_{23},q_{13},M_{S_j}^2,m_l^2,m_l^2,M_{V_i}^2)
\Big]
\\
&&
+2 M_{V_i}^2 
\Big[
D_0(0,0,0,M_H^2;q_{23},q_{12},M_{V_i}^2,M_{V_i}^2,m_l^2,M_{S_j}^2)
\n\\
&&\hspace{2cm}
- D_0(0,0,0,M_H^2;q_{23},q_{13},M_{V_i}^2,m_l^2,m_l^2,M_{S_j}^2)
\n\\
&&\hspace{2cm}
- (D_2 + D_0)(0,0,0,M_H^2;q_{12},q_{13},M_{V_i}^2,m_l^2,M_{S_j}^2,M_{S_j}^2)
\n\\
&&\hspace{2.cm}
- D_2(0,0,0,M_H^2;q_{12},q_{13},M_{S_j}^2,m_l^2,M_{V_i}^2,M_{V_i}^2)
\Big]
\n\\
&&
-3 \Big[ (C_{22}+C_{12}+C_2)(0,0,q_{13},m_l^2,m_l^2,M_{S_j}^2)
\n\\
&&\hspace{2cm}
+ (C_{22}+C_{12}+C_2)(0,0,q_{13},m_l^2,M_{S_j}^2,M_{S_j}^2)
\Big]
\n \\
&&\hspace{0cm}
+(C_{22}+C_{12}+C_2)(0,0,q_{13},m_l^2,m_l^2,M_{V_i}^2)
\n\\
&& \hspace{0cm}
-(C_{22}+C_{12}+C_2)(0,0,q_{13},m_l^2,M_{V_i}^2,M_{V_i}^2)
\Bigg\},
\n\\
F_{1,R}^{\text{Box}}|_{V_i,S_j}
&=& F_{1,L}^{\text{Box}}|_{V_i,S_j} 
(
g_{S_j l \nu_l}^R \rightarrow g_{S_j l \nu_l}^L;
g_{V_i l \nu_l}^L \rightarrow g_{V_i l \nu_l}^R 
),\\
F_{2,R}^{\text{Box}}|_{V_i,S_j} 
&=& F_{2,L}^{\text{Box}}|_{V_i,S_j}
(
g_{S_j l \nu_l}^R \rightarrow g_{S_j l \nu_l}^L;
g_{V_i l \nu_l}^L \rightarrow g_{V_i l \nu_l}^R 
). 
\end{eqnarray}
In the SM limit, we observe that these contributions are much smaller than 
other contributions because  of  the appearance of the factor $m_l$ in 
Eq.~(\ref{FLRBox-ViSj}). It means that we can take only the $\tau$-lepton 
contributions for these form factors. But in many BSMs, where new heavy 
charged leptons with $m_{E_l}\gg m_l$ appear in the loop. These contributions 
may be significant. For this case, the form factors 
are obtained directly by replacing $l$ by $E_l$. 
\begin{center}
\begin{figure}[ht]
\hspace{1.5cm}
\includegraphics[width=14.0cm, height=10.0cm]
{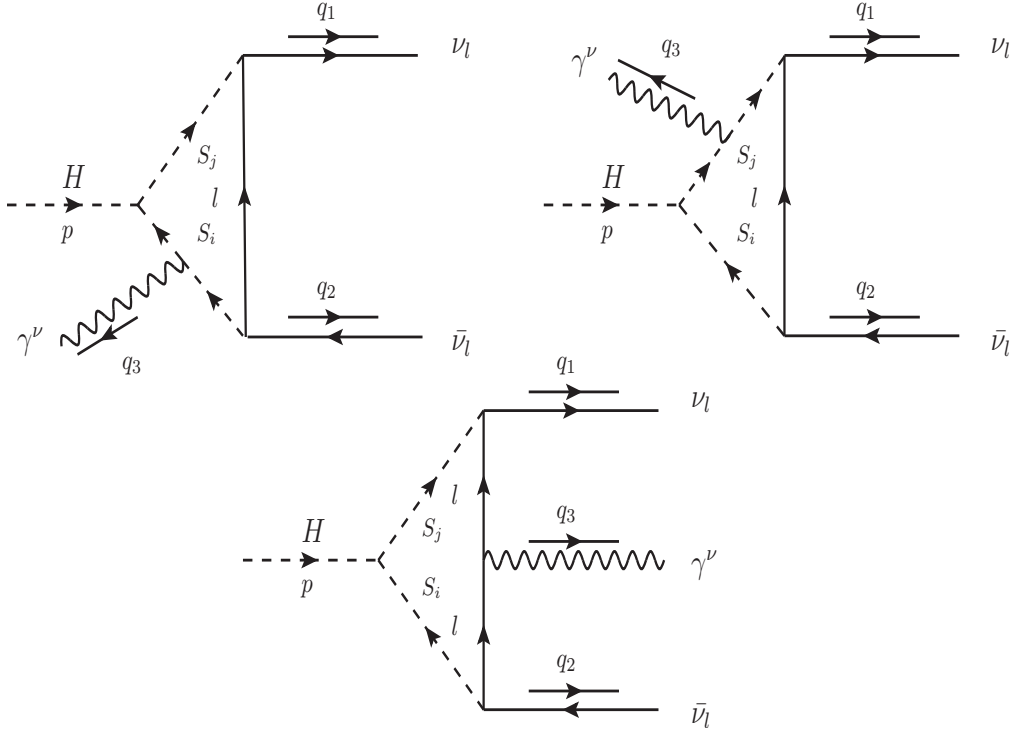}
\caption{Feynman box diagrams for $V_{i,j}, S_{i,j}$.}
\end{figure}
\end{center}
We finally end up with 
the contributions of one-loop box diagrams
with scalar charged bosons 
$S_i, S_j$ in the loop. The corresponding form 
factors read:
\begin{eqnarray}
\label{FLR}
F_{1,L}^{\text{Box}}|_{S_i,S_j}
&=& -\sum\limits_{S_i,S_j}
\dfrac{eQ  g_{HS_iS_j} \; g_{S_i l \nu_l}^R 
\; g_{S_j l \nu_l}^R }{8 \pi ^2}
\times
\n\\
&&
\times
\Bigg\{
(D_{33}
+D_{23}
+D_3)(0,0,0,M_H^2;q_{12},q_{13},M_{S_i}^2,m_l^2,M_{S_j}^2,M_{S_j}^2)
\n\\
&&
+(D_{33}
+D_{23}
+D_{13}
+D_3)(0,0,0,M_H^2;q_{23},q_{12},M_{S_i}^2,M_{S_i}^2,m_l^2,M_{S_j}^2)
\\
&&
+(D_{33}
+D_{23}+D_3)(0,0,0,M_H^2;q_{23},q_{13},M_{S_i}^2,m_l^2,m_l^2,M_{S_j}^2)
\Bigg\}, \n\\
F_{1,R}^{\text{Box}}|_{S_i,S_j}
&=& F_{1,L}^{\text{Box}}|_{S_i,S_j} 
( g_{S_i l \nu_l}^R \rightarrow g_{S_i l \nu_l}^L;
\; g_{S_j l \nu_l}^R  \rightarrow  g_{S_j l \nu_l}^L
),\\
F_{2,L}^{\text{Box}}|_{S_i,S_j}
&=& 
F_{1,L}^{\text{Box}}|_{S_i,S_j} 
( \{q_{13}, q_{23}\}
\rightarrow  \{q_{23}, q_{13}\}
),
\\
F_{2,R}^{\text{Box}}|_{S_i,S_j} 
&=& F_{2,L}^{\text{Box}}|_{S_i,S_j}
(g_{S_i l \nu_l}^R \rightarrow g_{S_i l \nu_l}^L;
\; g_{S_j l \nu_l}^R  \rightarrow  g_{S_j l \nu_l}^L). 
\end{eqnarray}
All the above form factors are checked numerically by
verifying ultraviolet finiteness of the results. 
We find that the results are good stability when 
varying ultraviolet cutoff parameters. We refer 
numerical results for this check in appendix $A$.

Having the correctness form factors for the
decay processes, the decay rate is given 
by~\cite{Kachanovich:2020xyg}: 
\begin{eqnarray}
\label{decayrate}
 \dfrac{d\Gamma}{d q_{12}q_{13}} 
 = \dfrac{q_{12}}{512 \pi^3 M_H^3}
 \Big[
 q_{13}^2(|F_{1,R}|^2 + |F_{2,R}|^2)
 + q_{23}^2(|F_{1,L}|^2 + |F_{2,L}|^2)
 \Big]. 
\end{eqnarray}
Taking the above integrand over $0 \leq q_{12} \leq M_H^2$ 
and $0\leq q_{13} \leq M_H^2 - q_{12}$, one gets the 
total decay rate. In the next subsection, we show a 
typical example which we apply the analytic results 
for $H\rightarrow \nu_l \bar{\nu}_l \gamma$ in standard 
model. Phenomenological results for these decay channels
also studied with using updated parameters at the LHC. 
\subsubsection{Standard model case}
In this case, we have $V_i, V_j \rightarrow W^+, W^-$, 
$V_k^0 \rightarrow Z$. All couplings are replaced 
by $g_{HV_iV_j} = eM_W/s_W,\; g_{V^0_kV_iV_j} = e \; c_W/s_W, 
\; g_{V^0_kAV_iV_j} = e^2 \; c_W/s_W, 
\; g_{V^0_kNN}^L = e/(2 s_W c_W),\; g_{V^0_kNN}^R = 0, \;
g_{Hf_if_j}^L= g_{Hf_if_j}^R  = e \; m_f/(2 s_W\; M_W),\; 
g_{V^0_kf_if_j}^L = e(T_{3}^f - Q_f\; s^2_W)/(s_W \; c_W), 
\; g_{V^0_kf_if_j}^R = -e Q_f s_W/c_W, 
\; g_{V_iNl}^L = e/(\sqrt{2}\;s_W), 
\; g_{V_iNl}^R = 0$. Analytic results  
for case of $m_l\rightarrow 0$ are presented as follows:
\begin{eqnarray}
F_{1,L}^{\text{Trig, SM}}|_{W,W} 
&=&
\dfrac{\alpha^2}{4 M_W^3 s_W^3
(q^2 - M_{Z}^2 + i\Gamma_{Z}M_{Z})}
\times
\\
&&
\times
\Bigg\{
\Big[
M_H^2 (2 B_{111}
+3 B_{11}
+ B_1)
+2 B_{00}
+4 B_{001}
\Big](M_H^2,M_W^2,M_W^2)
\n \\
&&
+\Big[
4 M_H^2 M_W^2
-2 M_H^2 q^2
+8 (d-1) M_W^4
-4 M_W^2 q^2
\Big] 
\times
\n \\
&&\hspace{3.0cm}
\times
\Big[
C_{22}+C_{12}+C_2
\Big]
(0,q^2,M_H^2,M_W^2,M_W^2,M_W^2)
\n \\
&&
+4 M_W^2 (4 M_W^2-q^2) 
C_0(0,q^2,M_H^2,M_W^2,M_W^2,M_W^2)
\Bigg\},
\n \\
F_{1,L}^{\text{Trig, SM}}|_{f,f} 
&=&
-\dfrac{\alpha^2 m_f^2 N_C^f Q_f }{2 c_{W}^2 s_{W}^3 M_{W}
(q^2 - M_{Z}^2 + i\Gamma_{Z}M_{Z})}
\left(2 Q_f s_{W}^2-T_3^{f}\right)
\times
\\
&&
\times
\Bigg\{
C_0(0,q^2,M_H^2,m_f^2,m_f^2,m_f^2)
+4
\Big[
C_{22}+C_{12}+C_2
\Big](0,q^2,M_H^2,m_f^2,m_f^2,m_f^2)
\Bigg\}.
\n
\end{eqnarray}
We also have $F_{k,R}^{\text{Trig, SM}}|_{W,W}
= F_{k,R}^{\text{Trig, SM}}|_{f,f} = 0$ for $k=1,2$
due to the fact that all couplings $g^R_{\cdots}$ 
are absent in SM. For one-loop box diagrams, 
the form factors read
\begin{eqnarray}
\label{FLR}
F_{1,L}^{\text{Box, SM}}|_{W,W} &=& 
\dfrac{\alpha^2}{2 M_W^3 s_W^3}
\Bigg\{\; (M_H^2 +2 M_W^2)
\Big[
(C_{22} +C_{12} +C_2)(0,q_{12},M_H^2,M_W^2,M_W^2,M_W^2)
\n \\
&&
\hspace{4.5cm}
+(C_{22} +C_{12} +C_2)(q_{12},0,M_H^2,M_W^2,M_W^2,M_W^2)
\Big]
\n \\
&&
+ 2 (d-2) M_W^4
\Big[
(D_{33} +D_{23})(0,0,0,M_H^2;q_{12},q_{13};M_W^2,0,M_W^2,M_W^2)
\n \\
&&
\hspace{2.5cm}
+(D_{33} +D_{23})(0,0,0,M_H^2;q_{23},q_{13};M_W^2,0,0,M_W^2)
\n \\
&&\hspace{2.5cm}
+ (D_{33} +D_{23} 
+ D_{13})(0,0,0,M_H^2;q_{23},q_{12};M_W^2,M_W^2,0,M_W^2)
\Big]
\n \\
&&
+ 2 (d-4) M_W^4
\Big[D_3(0,0,0,M_H^2;q_{12},q_{13};M_W^2,0,M_W^2,M_W^2)
\n \\
&&\hspace{3cm}
+D_3(0,0,0,M_H^2;q_{23},q_{12};M_W^2,M_W^2,0,M_W^2)
\n \\
&&\hspace{3cm}
+D_3(0,0,0,M_H^2;q_{23},q_{13};M_W^2,0,0,M_W^2)
\Big]
\n \\
&&
+ 4 M_W^2 
\Big[
C_0(0,q_{12},M_H^2,M_W^2,M_W^2,M_W^2)
\\
&&\hspace{4cm}
- M_W^2 D_2(0,0,0,M_H^2;q_{12},q_{13};M_W^2,0,M_W^2,M_W^2)
\Big]
\Bigg\},
\n\\
F_{2,L}^{\text{Box, SM}}|_{W,W} &=& 
F_{1,L}^{\text{Box, SM}}|_{W,W}
\left(\{ q_{13}, q_{23}\}\rightarrow
\{q_{23}, q_{13} \} \right), \\
F_{1,R}^{\text{Box, SM}}|_{W,W} &=&
F_{2,R}^{\text{Box, SM}}|_{W,W}
= 0.
\end{eqnarray}
For phenomenological results, 
we use following input parameters:
$M_Z = 91.1876$ GeV, 
$\Gamma_Z  = 2.4952$ GeV, $M_W = 80.379$ GeV, $M_H =125.1$ GeV,
$m_{\tau} = 1.77686$ GeV, $m_t= 172.76$ GeV, 
$m_b= 4.18$ GeV, $m_s = 0.93$ GeV and  $m_c = 1.27$ GeV. 
We first confirm the previous result in Ref.~\cite{Sun:2013cba} 
which the decay rate is computed in $\alpha$-scheme, or 
$\alpha = 1/137.035 999 084$. By working in this scheme, 
the decay rate (for $l=e$) is obtained as
$\Gamma_{H\rightarrow \nu_e\bar{\nu}_e\gamma} 
= 0.480414\; \text{KeV}$.
This value gives a good agreement with the result in 
Ref.~\cite{Sun:2013cba}.

\begin{figure}[ht]
\begin{center}$
\begin{array}{lr}
\dfrac{d\Gamma}{dm_{\nu\bar{\nu}}} & \\
\includegraphics[width=15cm, height=6cm]
{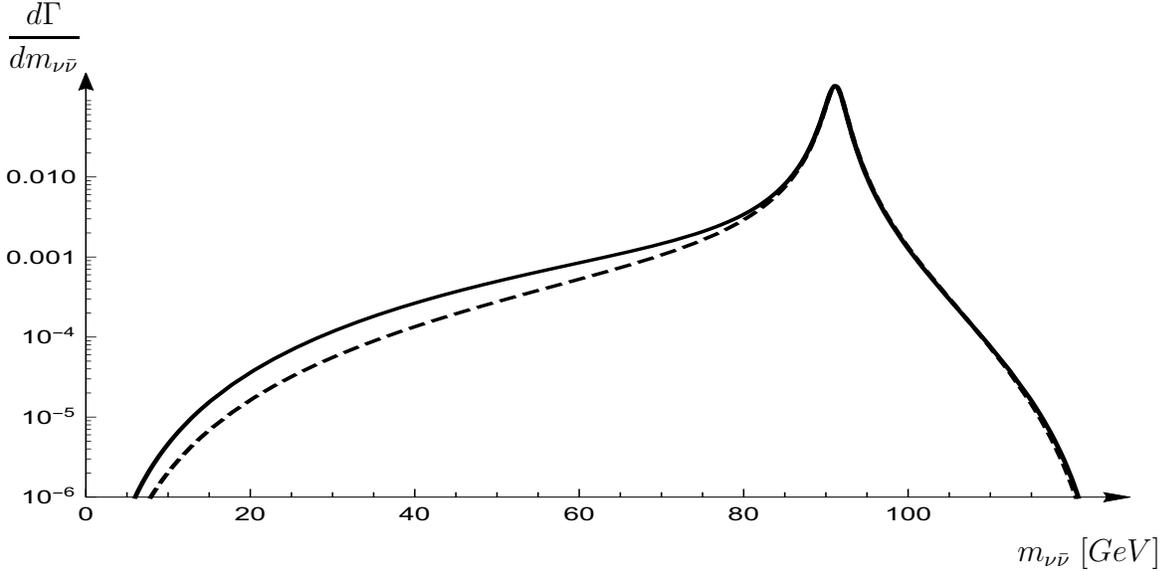}\\
&\hspace{-2cm}m_{\nu\bar{\nu}}\; [GeV]
\end{array}$
\end{center}
\caption{\label{mNuNu} Differential decay rate is plotted 
as function of invariant mass of $m_{\nu_l\bar{\nu}_l}$.}
\end{figure}

At the LHC, the decay processes are involved two kind of events: 
(i) considering photon is undetected we then have Higgs decay to
invisible particles; (ii) for detected photon, we observe 
the Higgs decay to photon plus missing energy. The former events
provide important information for controlling SM background
for $H\rightarrow \gamma\gamma$ and $H\rightarrow Z\gamma$ 
which $Z$ may decay to undetected leptons, etc. For latter events, 
they are interesting for searching dark matter at the LHC. 
For above reasons, both the events are examined in this 
paper with using updated parameters at the LHC. In this 
computation, we work in $G_F$-scheme 
which $\alpha$ is evaluated from 
$G_F = 1.1663787\times 10^{-5}$ GeV$^{-2}$. 
The resulting reads
\begin{eqnarray}
\alpha^{-1} = \dfrac{\pi}{\sqrt{2}G_F M_W^2 s_W^2} = 132.184.  
\end{eqnarray}
The new results for decay rates are obtained: 
\begin{eqnarray}
\Gamma^{\text{Trig}}_{H\rightarrow \nu_l\bar{\nu}_l\gamma}  
=  0.536234\; \text{KeV},\\
\Gamma^{\text{Tot}}_{H\rightarrow \nu_l\bar{\nu}_l\gamma} 
= 0.554933\; \text{KeV}.
\end{eqnarray}
We realize that the attributions of $|F_{k/L(R)}^{\text{Box}}|^2$ 
is much smaller than the other terms.  Differential decay rate is also 
plotted as function of invariant mass of 
$m_{\nu_l\bar{\nu}_l}$(or $m_{\nu_l\bar{\nu}_l}=\sqrt{q_{12}})$. 
The distribution is defined in form of:
\begin{eqnarray}
\dfrac{d\Gamma}{d m_{\nu_l\bar{\nu}_l}} 
= \dfrac{m_{\nu_l\bar{\nu}_l}^3 }{512 \pi^3 M_H^3} 
\int\limits_{0}^{M_H^2 -2 m_{\nu_l\bar{\nu}_l}^2 } 
\; d q_{13} 
\Big[
 q_{13}^2(|F_{1,R}|^2 + |F_{2,R}|^2)
 + q_{23}^2(|F_{1,L}|^2 + |F_{2,L}|^2)
 \Big]. 
\end{eqnarray}
The distribution is shown in Fig.~\ref{mNuNu}. We observe 
a peak of $Z^*\rightarrow \nu_l\bar{\nu}_l$ which is around 
$M_Z$. In the region $m_{\nu\bar{\nu}} \leq M_Z$, the contributions
of box diagrams are visible. While they give a small contribution
beyond the peak. 

We are also interested in the case of photon that can be tested at 
the colliders. In this case, one should apply the energy cuts 
for final photon. The results are shown with different cuts for
photon in Table~\ref{testPHOTON}. The updated results are important
should take into account at the HL-LHC and future colliders.
\begin{table}[h]
\begin{center}
\begin{tabular}{c|ccc}  \hline \\
$\Gamma$ [KeV]/$E_{\gamma}^{\text{cut}}$ [GeV]
& $5$    & $30$      & $50$ \\ \hline\hline \\
$\Gamma_{H\rightarrow \gamma \nu_l \bar{\nu}_l}^{\text{Trig}} $
&$ 0.536232$ & $0.188952$ & $ 0.00529925$\\ \hline \\
$\Gamma_{H\rightarrow \gamma \nu_l \bar{\nu}_l}^{\text{Total}} $
&$0.554931$ & $0.20927$ & $0.00993683$ \\ \hline\hline 
\end{tabular}
\caption{\label{testPHOTON} Decay widths in the case of photon 
that can be tested.}
\end{center}
\end{table}

We note that all numerical results shown in this subsection 
are for a family of neutrino in final state. For all neutrinos, 
we multiply factor $3$ for all above results. 
\section{Conclusions}   
We have presented analytic formulas for all possible 
one-loop contributions to the SM-like Higgs decay 
$H\rightarrow \nu_l\bar{\nu}_l\gamma$  that are valid 
in many BSMs. Additional vector bosons, charged fermions and 
charged (and also neutral) scalar 
particles exchanging in the loop diagrams 
have considered in this computation. 
General statement, we conclude that the 
evaluations can be extended directly for 
general numbers of the extra neutrinos in final 
states. Analytic results are expressed in a general form 
which written in terms of Passarino-Veltman scalar 
functions that can be evaluated numerically 
using {\tt LoopTools}. The computations have  
checked numerically by verifying ultraviolet finiteness 
of the results. We find that the results are good 
stability when varying ultraviolet cutoff parameters. 
We then apply the results to the standard model which  
the decay rates are generated and cross-checked to 
previous computation. All physical results for the decay 
channels within standard model are studied with 
the updated input parameters at the 
Large Hadron Collider.  
\\

\noindent
{\bf Acknowledgment:}~
This research is funded by Vietnam National Foundation
for Science and Technology Development (NAFOSTED) under 
the grant number No.103.01-2019.387.\\
\appendix 
\section{Numerical checks for the calculations}       
Numerical checks for the computations
are performed for all the above form factors. 
The results must be independent of ultraviolet cutoff 
($C_{UV}=1/\varepsilon$) and 
$\mu^2$ parameters. For demonstrating, we take 
the form factors $F_{1L}^{\text{Trig}}|_{WW}$ 
and $F_{1L}^{\text{Box}}|_{WW}$, appear
high rank tensor one-loop integrals in the amplitude, 
as typical examples. Numerical results are presented 
at arbitrary sampling point in physical region.
\begin{table}[h]
\begin{center}
\begin{tabular}{l|ccc}  \hline \\
$\textbf{Diagrams} /(C_{UV}, \mu^2)$
& $(0, 1)$ & $(10^{5}, 10^{7})$   \\ \\ \hline \hline\\
$1st$  & $5.684478386592405 \cdot 10^{-8}$ & 
$-0.0004193697635384515$ \\ 
& + $7.556282593901243 \cdot 10^{-8} \, i$ &  
$-0.0005574612531042949 \, i$  \\ \hline\\
$2nd$  & $5.684478386592405 \cdot 10^{-8}$ & $-0.0004193697635384515$ \\ 
& + $7.556282593901243 \cdot 10^{-8} \, i$ &  $-0.0005574612531042948 \, i$  
\\ \hline\\
$3rd$  & $-6.951714952517838 \cdot 10^{-8}$ & $0.00083878369949511$  \\ 
&  $-9.24079908850924 \cdot 10^{-8} \, i$ & + $0.0011149812238695827 \, i$  
\\ \hline\\
Sum & $4.417241820666968\cdot 10^{-8}$ & $4.417241820666968\cdot 10^{-8}$  \\   
 & $+5.871766099293248 \cdot 10^{-8} \, i$ & $+5.871766099293248 \cdot 10^{-8} \, i$ 
 \\
 \hline\hline
\end{tabular}
\caption{Numerical checks for 
$F_{1L}^{\text{Trig}}|_{WW}$.}
\end{center}
\end{table}
\begin{table}[h]
\begin{center}
\begin{tabular}{l|ccc}  \hline \\
$\textbf{Diagrams} /(C_{UV}, \mu^2)$
& $(0, 1)$ & $(10^5, 10^7)$   \\ \\ \hline \hline\\
$1st$  & $-3.114167099931247 \cdot 10^{-10}$ & $-3.114167099931247 \cdot 10^{-10}$ \\ \hline\\
$2nd$  & $6.440660243424821 \cdot 10^{-10}$ & $6.440660243424821 \cdot 10^{-10}$ \\  \hline\\
$3rd$  & $-9.02406987251144 \cdot 10^{-11}$ & $-9.02406987251144 \cdot 10^{-11}$ \\ \hline\\
Sum & $2.424086156242413 \cdot 10^{-10}$ & $2.424086156242413 \cdot 10^{-10}$  \\   
   \hline\hline
\end{tabular}
\caption{Numerical checks for $F_{1L}^{\text{Box}}|_{WW}$.}
\end{center}
\end{table}
\section{Feynman rules}       
\begin{center}
\begin{table}[h!]
\centering
{\begin{tabular}{l@{\hspace{2cm}}l }
\hline \hline
\textbf{Particle types} & \textbf{Propagators}\\
\hline \hline \\
Fermions $f$ 
& $i\dfrac{\slashed{k} + m_{f}}{k^2-m_{f}^2}$ \\
Gauge boson $V_i$
& $\dfrac{- i}{p^2 - M_{V_i}^2} 
\Bigg( g^{\mu \nu} - \dfrac{p^\mu p^\nu}{M_{V_i}^2}  \Bigg)$ \\ 
Gauge boson $V^{0}_k$
& $\dfrac{- i}{ p^2 - M_{V_k^{0}}^2 + i\Gamma_{ V^{0}_{k} } M_{V_k^{0}} } 
\Bigg( g^{\mu \nu} - \dfrac{p^\mu p^\nu}{M_{V^{0}_k}^2}  \Bigg)$ \\ 
Charged (neutral) scalar bosons $S_i(S_k^0)$
& $\dfrac{i}{p^2 - M_{S_i}^2 (M_{S^0_k}^2)}$
\\  \\
\hline \hline \\
\end{tabular}}
\caption{Feynman rules involving the decay
in unitary gauge.
\label{Feynman rules table}}
\end{table}
\end{center}
\begin{table}[h!]
\begin{center}
\begin{tabular}{l@{\hspace{2cm}}l} 
\hline \hline
\textbf{Vertices} & \textbf{Couplings}   \\ \hline \hline
\\
$H \cdot \bar{f_i}  \cdot f_j $ 
& $-i \, \Big( g_{Hf_if_j}^{L}
P_L + g_{Hf_if_j}^{R} P_R \Big)$
\\ \\
$H \cdot V_i^\mu \cdot V_j^\nu $ 
& $i \, g_{HV_iV_j} \, g^{\mu \nu}$ 
\\ \\
$H \cdot S_i  \cdot S_j $
& $-i \, g_{HS_iS_j} $
\\ \\
$H (p) \cdot V_i^\mu \cdot S_j (q)$
& $ i \, g_{HV_iS_j} \, (p-q)^\mu$
\\ \\
$A^\mu \cdot f_i \cdot \bar{f_i} $
& $i e Q_f \gamma^\mu $ 
\\ \\
$A^\mu (p_1) \cdot V^{Q\nu}_i  (p_2) \cdot V^{-Q\lambda}_i  (p_3)$
&$-i e Q\, \Gamma^{\mu \nu \lambda} (p_1, p_2, p_3)$ 
\\ \\
$A^\mu \cdot S^{Q}_i (p) \cdot S^{-Q}_i (q)$
& $i e Q\, (p-q)^\mu$
\\ \\
$V^{0 \; \mu}_k \cdot f_i \cdot \bar{f_j}$
& 
$i \gamma^\mu \Big( g_{V^0_k f_if_j}^{L} P_L 
+ g_{V^0_k f_if_j}^{R} P_R \Big)$
\\ \\
$V^{0 \; \mu}_k (p_1) \cdot V_i^\nu  (p_2) \cdot V_j^\lambda  (p_3)$
&
$-i \, g_{V^0_k V_iV_j}\; \Gamma^{\mu \nu \lambda} (p_1, p_2, p_3)$ 
\\ \\
$V^{0 \; \mu}_k \cdot S_i  (p) \cdot S_j  (q)$
& 
$i \, g_{V^0_k S_iS_j} \, (p-q)^\mu$   
\\ \\
$V^{0 \; \mu}_k \cdot V_i^\nu  \cdot S_j $
& 
$ g_{V^0_k V_iS_j} \, g^{\mu \nu}$ 
\\ \\
$V^{\mu}_i \cdot \bar{l} \cdot \nu_l $  &
$i \gamma^\mu \Big( g_{V_i l \nu_l}^L P_L + g_{V_i l \nu_l}^R P_R \Big) $
\\ \\
$S_i \cdot \bar{l} \cdot \nu_l $  &
$i g_{S_i l \nu_l}^{L} P_L + i g_{S_i l \nu_l}^{R} P_R $
\\ \\
$S^*_i \cdot l \cdot \bar{\nu}_l $  &
$i g_{S_i l \nu_l}^{R} P_L + i g_{S_i l \nu_l}^{L} P_R $
\\ \\
$V^{0 \; \mu}_k \cdot A^\nu \cdot V_i^\alpha \cdot V_j^\beta$
& 
$-i \, g_{V^0_k AV_iV_j} \, S^{\mu \nu, \alpha \beta}$
\\ \\
\hline \hline
\end{tabular}
\end{center}
\caption{Feynman couplings  in the unitary gauge 
with $P_{L/R} = (1 \mp \gamma_5)/2$,
$\Gamma^{\mu \nu \lambda} (p_1, p_2, p_3) 
= g^{\mu \nu} (p_1 - p_2)^\lambda + g^{\lambda \nu}
(p_2 - p_3)^\mu + g^{\mu \lambda} (p_3 - p_1)^\nu$ 
and $S^{\mu \nu, \alpha \beta} 
= 2 g^{\mu \nu} g^{\alpha \beta} - g^{\mu \alpha} g^{\nu \beta} 
- g^{\mu \beta} g^{\nu \alpha}$ 
and $Q$ denotes the electric charge of the gauge and 
charged bosons $V^{Q}_i$ and $S^Q$. When we consider the extra
charged leptons $E_l$ in the loop,  the results will be obtained directly
by replacing $l$ by $E_l$ respectively. 
\label{couplings table}}
\end{table}

\end{document}